\documentclass[12pt,preprint]{aastex}

\begin{document}
\title{Cyclical Changes in the Timing Residuals from
      the Pulsar B0919+06}
\author{Tatiana V. Shabanova\altaffilmark{}}
\affil{Pushchino Radio Astronomy Observatory, Astro Space Center,
 P. N. Lebedev Physical Institute, Russian Academy of Sciences,
 142290 Pushchino, Russia}
\email{tvsh@prao.ru}

\begin{abstract}
We report the detection of a large glitch in the pulsar B0919+06
(J0922+0638). The glitch occurred in 2009 November 5 (MJD 55140)
and was characterized by a fractional increase in the rotation
frequency of $\Delta\nu/\nu \sim 1.3\times 10^{-6}$. A large
glitch happens in the pulsar whose rotation has unstable
character. We present the results of the analysis of the rotation
behavior of this pulsar over the 30-year time span from 1979 to
2009. These results show that the pulsar's rotation frequency
underwent continuous, slow oscillations which look like glitch-like
events. During the 1991--2009 interval, the pulsar experienced a
continuous sequence of 12 slow glitches with a fractional increase
in the rotation frequency $\Delta\nu/\nu \sim 1.5 \times10^{-9}$.
All the slow glitches observed have a similar signature related to
a slow increase in the rotation frequency during $\sim$ 200 days
and the subsequent relaxation back to the pre-glitch value during
$\sim$ 400 days. We show that a continuous sequence of such slow
glitches is characterized by practically identical amplitudes
$\Delta\nu \sim 3.5 \times10^{-9}$ Hz and identical time intervals
between glitches $\sim$ 600 days and is well described by a
periodic sawtooth-like function. The detection of two different
phenomena, such as a large glitch and a sequence of slow glitches,
indicates the presence of two types of discontinuities in the
rotation frequency of the pulsar B0919+06. These discontinuities
can be classified as normal and slow glitches.
\end{abstract}


\section{Introduction}

The pulsar B0919+06 was discovered in the second Molonglo pulsar
survey \citep{man78}. It has a period of 0.430 s,
a period derivative of $13.72\times10^{-15}$, and
a characteristic age of $\tau={P}/{2\dot{P}} \sim 5\times10^{5}$ years.
A pulsar distance of 1200 pc and a transverse speed of 505 km s$^{-1}$
were derived from the measurements of astrometric parameters
\citep{cha01}.

The data set analyzed includes the archival Jet Propulsion
Laboratory (JPL) data and the Pushchino Radio Astronomy
Observatory (PRAO) data. The early timing observations of PSR
B0919+06 were carried out at the JPL at the frequency of 2388 MHz
between 1979 December and 1983 March and covered a span of 3.2
years \citep{dow86}. At the PRAO, timing observations of the
pulsar were made between 1983 August and 2010 February. Together,
both the data sets cover the 30-year span from 1979 to 2009 with a
four-year gap between 1987 and 1990. An analysis of these
experimental data showed that PSR B0919+06 has unstable rotation
over the entire data span. The timing residuals after the removal
of deterministic pulsar spin-down from the arrival times are
characterized by a large second derivative that indicates a high
level of timing noise.

Two types of unpredictable variations may occur in the spin rate
of pulsars -- timing noise and glitches \citep{she96}.
Timing noise is manifestation of random, continuous wandering of
the pulse phase that may produce long-term polynomial trends in
the timing residuals. Glitches represent sudden, discrete jumps in
the pulsar's rotation frequency, followed by an exponential
recovery to the pre-glitch value. Glitches are characterized by
short rise times of less than one day and reveal themselves as sudden
discontinuities in the timing residuals.

A study of the timing behavior of the pulsars B1822$-$09 and B1642$-$03
has shown that there is another type of glitches which can be classified
as peculiar or slow glitches
\citep[hereafter SH09]{sha98,sha00,zou04,sha05,sha07,sha09a,sha09b}.
Characteristic features of slow glitches are long rise times of about
200--500 days and small amplitudes of about several parts in $10^{-9}$ Hz.
Slow glitches produce cyclical changes in the timing residuals.
The rotation frequency of these pulsars undergoes continuous, slow
oscillations during a long period of time.
For the pulsar B1822$-$09, the oscillatory changes in the rotation
frequency were observed over the 1995--2004 interval. In the case of
PSR B1642$-$03, cyclical changes in the timing residuals are observed
during the 40-year period from 1969 to 2008. This pulsar
demonstrates such striking properties of the slow glitches which
allow us to predict the epochs and the amplitudes of new glitches
in its rotation frequency (see SH09).

In this paper, we study the rotation history of the pulsar B0919+06
over the 30-year period, report the detection of a large glitch, and
show that cyclical changes in the timing residuals from this pulsar
are due to the presence of slow glitches. These results indicate
that two types of discontinuities, specified as normal and slow glitches,
can occur in the rotation frequency of one pulsar.

\section{Observations and Timing Analysis}

Timing observations of the pulsar B0919+06 were carried out at
the Pushchino Observatory for more than 26 years from 1983 August
to 2010 February with a four-year gap between 1987 and 1990.
The observations were made with the Large Phased Array of the
Pushchino Observatory, which is a transit telescope, at frequencies
near 102 or 112 MHz using a 64-channel radiometer with a channel
bandwidth of 20 kHz and a time resolution of 2.56 or 1.28 ms.
The duration of one observation of PSR B0919+06 was determined
by the width of the antenna beam at the pulsar declination and
lasted 3.2 minutes. During this time, 450 individual pulses were
summed synchronously with a predicted topocentric pulsar period
to produce the mean pulse profile in each 20 kHz channel.
After dispersion removal, all the channel profiles were summed
to form an integrated pulse profile for the given observing session.
Then this integrated profile was cross-correlated with a standard
low-noise template to give the topocentric pulse arrival times
for each observing session.

The topocentric arrival times collected at PRAO and the
geocentric arrival times obtained from the archival JPL timing
data were all referred to as the barycenter of the solar system at
infinite frequency using the program TEMPO \footnote
{http://www.atnf.csiro.au/research/pulsar/tempo} and the JPL DE200
ephemeris. The position and the proper motion that are required for
this correction were taken from \citet{hob04} and \citet{cha01},
respectively.
In order to obtain residuals from the timing model, the pulsar's
rotation is modeled by a polynomial including several frequency
derivatives. In accordance with a Taylor expansion, the pulse phase
$\varphi$ at the barycentric arrival time $t$ is expressed as:
\begin{equation}
\varphi(t) = \varphi_{0} + \nu(t-t_{0}) + {\dot\nu}(t-t_{0})^{2}/2 +
{\ddot\nu}(t-t_{0})^{3}/6...,
\end{equation}
where $\varphi_{0}$, $\nu$, $\dot{\nu}$, and $\ddot{\nu}$ are the
pulse phase, rotation frequency, first and second frequency
derivative at some reference time $t_{0}$, respectively.
The timing residuals, obtained as differences between the observed
times and the times predicted by a best fit model, were used
for an analysis of the rotation behavior of the pulsar.
The pulsar position was held fixed in the fitting procedure.

In order to study variations in the pulsar's rotation,
the parameters $\nu$ and $\dot\nu$ were calculated from the local
fits, performed to the pulse arrival times over the intervals of 150
or 300 days. The frequency residuals $\Delta\nu$ were obtained at
the initial epoch of each interval relative to a third-order
polynomial (1), including the mean parameters $\nu$, $\dot\nu$, and
$\ddot\nu$ defined over the full interval 1979--2009, 1991--2009
or any other interval.

\section{Results}

\subsection{A Large Glitch of 2009}

\begin{figure}
\epsscale{.80}
\plotone{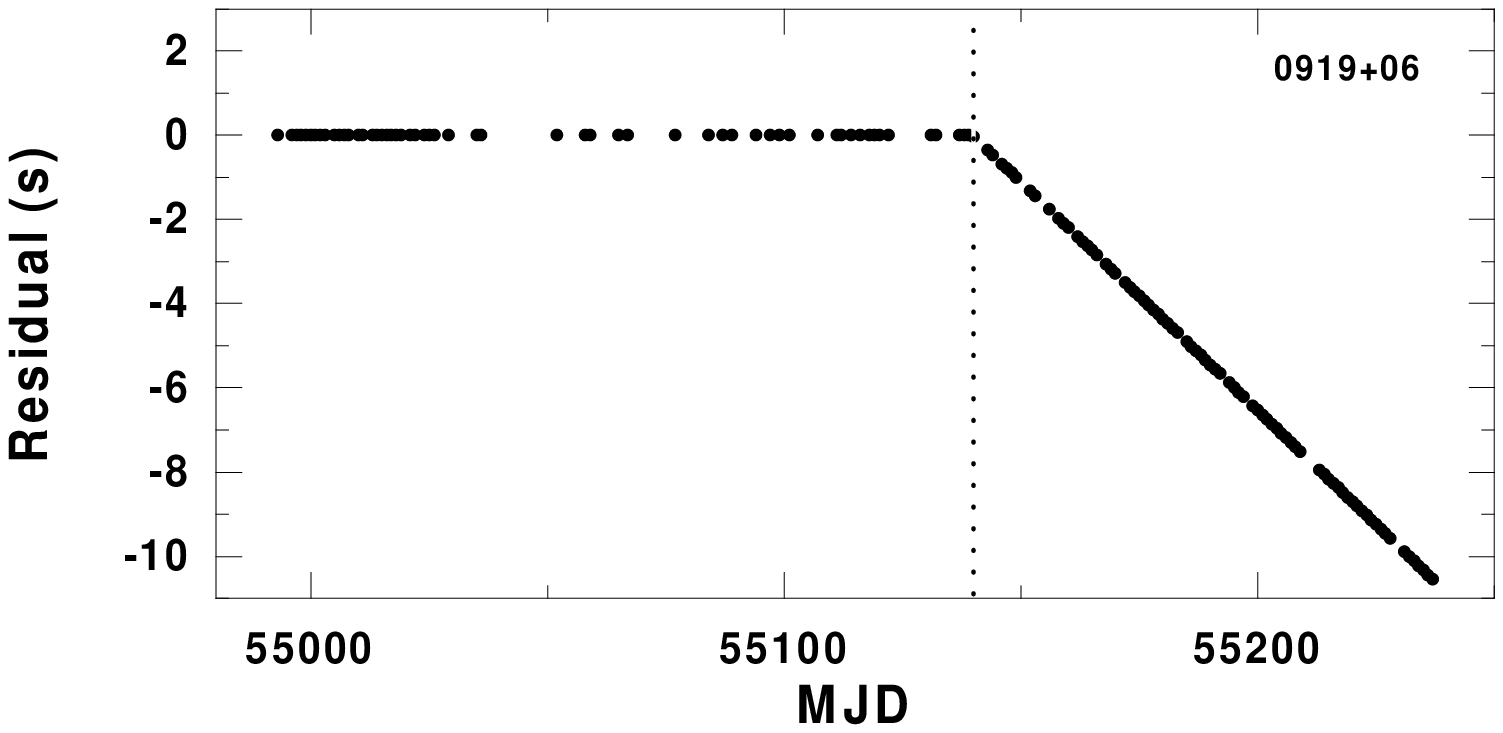}
 \caption{Timing residuals for the large glitch that occurred
      in the period of PSR B0919+06 in 2009 November 5 (MJD 55140).
      The dotted line indicates the glitch epoch.
     \label{glitch}}
\end{figure}

In 2009 November 5, the pulsar B0919+06 suffered a very large glitch
with the fractional increase in the rotation frequency
$\Delta\nu/{\nu}=1.3 \times 10^{-6}$. This glitch is comparable in size
to the glitches observed in the Vela pulsar. The glitch was accompanied
by a significant decrease in the frequency derivative
${\Delta{\dot\nu}}/{\dot\nu}=-7 \times 10^{-3}$.
A glitch was detected during a series of daily observations,
so the uncertainty in determining the glitch epoch was within several
hours, MJD 55139.8(1). Figure~\ref{glitch} shows the timing residuals
relative to a simple $\nu$, $\dot\nu$ model fitted to the data before
the glitch. The negative growth in the residuals due to this glitch
corresponds to a shift of the pulse in the observing window by
$\sim$ 109 ms/day.
The glitch parameters are given in Table~\ref{glitchp}. Uncertainties
in parentheses represent the formal standard deviation and refer
to the last digit quoted.

The large glitch has occurred in a relatively old pulsar
with a characteristic age of $\tau\sim 5\times10^{5}$ years. This age
is comparable to the age of PSR B0355+54
($\tau\sim 5.6\times10^{5}$ years) that experienced a giant glitch
with $\Delta\nu/{\nu}=4.4 \times 10^{-6}$ in 1987 \citep{lyn87}.
The timing observations of PSR B0919+06 over four months after
the glitch are not sufficient for the detailed research of the
post-glitch behavior. Moreover, as will be shown in the following sections,
the rotation frequency of this pulsar undergoes continuous, slow
oscillations with the spacing of maxima $\sim$ 600 days. Further
observations are needed to establish a relationship between
the large glitch and this oscillatory behavior.

\subsection{The Timing Behavior of the Pulsar Between 1979 and 2009}

The timing behavior of PSR B0919+06 over the 30-year interval from
1979 December to 2009 November before the large glitch occurred
is presented in Figure~\ref{resid1}. A four-year data
gap seen in the residuals between 1987 and 1990 insignificantly
distorts the information about the pulsar's rotation.

An analysis of the full data set showed that the timing behavior of
PSR B0919+06 exhibits significant deviations from a deterministic
spin-down low that indicates a presence of large timing noise.
So, the pulsar's rotation was modeled by polynomials including several
frequency derivatives.
The timing residuals after subtraction of the third-order polynomial
for $\nu$ and two frequency derivatives $\dot\nu$ and $\ddot\nu$
are shown in Figure~\ref{resid1}(a). The corresponding rotation
parameters for the model 1979--2009 are given in Table~\ref{model}.
The post-fit residuals display a large quartic term with
the amplitude approximately equal to half the pulsar period. This
plot shows that the residual curve has weak but well noticeable
short-term cyclical structure over the entire time span.

This cyclical structure become more discernible in the timing
residuals presented in Figure~\ref{resid1}(b). The post-fit
residuals obtained after subtraction of a polynomial including $\nu$
and three frequency derivatives exhibit the three maxima of
a large-scale structure. The timing model with higher-order
derivatives, for example, with four or five frequency derivatives,
gives similar post-fit residuals and does not remove this
structure from the timing residuals. Its origin is uncertain.
We pay attention to the short-term cyclical structure
superimposed on this residual curve.
It contains 19 cycles which are located through the intervals
approximately equal to 600 days.
The properties of just this cyclical structure will be studied
in detail in the following sections.

The changes in the rotation frequency with time are shown in
Figure~\ref{resid1}(c). The curve A marks the frequency residuals
$\Delta\nu$ relative to the mean rotation parameters
$\nu$, $\dot\nu$, and $\ddot\nu$  over
the 1979--2009 interval (Table~\ref{model}).
The values of $\nu$ were calculated from the local fits, performed
to the pulse arrival times over the intervals of $\sim$ 300 days.
This interval is approximately equal to half a cycle duration
in the timing residuals. The strong curvature of the $\Delta\nu$
curve is due to the presence of higher-order frequency
derivatives which are not taken into account by the timing model.

The curve A clearly shows that the signature of the frequency
residuals $\Delta\nu$ has a sawtooth-like character. This pattern
is well appreciable over the interval 1991--2009. The observed
appearance of the $\Delta\nu$ changes indicates that the rotation
frequency of the pulsar underwent slow oscillations during all
the observational interval. These slow oscillations in $\nu$
give rise to cyclical changes in the timing residuals.
Our purpose is to determine the cause of slow oscillations in
the pulsar's rotation frequency. Because of the paucity of data between
1979 and 1986, the properties of this oscillatory structure will be
investigated over the time span from 1991 to 2009 where there is
a great deal of experimental data.

\subsection{Slow Oscillations in the Pulsar's Rotation Frequency
            over the Interval 1991--2009}

Figure~\ref{resid2} exhibits the timing behavior of the pulsar over
the 19-year period between 1991 January and 2009 November.
Figure~\ref{resid2}(a) displays the timing residuals after subtraction
of a polynomial including $\nu$ and two frequency derivatives.
The rotation parameters measured over this period are given
in Table~\ref{model}. The last point in the residual curve indicates
the large glitch that occurred in 2009 November 5. The short-term
cyclical structure is clearly seen in the timing residuals.

The timing residuals after subtraction of a polynomial including $\nu$
and five frequency derivatives are given in Figure~\ref{resid2}(b).
This plot shows a clear cyclical structure that contains 12 cycles
with the amplitudes of about 10 ms and the spacing of maxima
of about 600 days. Comparison of the timing residuals presented in
Figures~\ref{resid1}(a) and \ref{resid1}(b) and
Figures~\ref{resid2}(a) and \ref{resid2}(b) exhibits that the epochs
and the spacing of the maxima of these cycles are the same
in all the plots and do not
depend on the time span of the data analyzed and the polynomial model
fitted. At the same time, the shape and amplitude of these cycles
are different in these plots.

The time behavior of the frequency residuals $\Delta\nu$ and
the frequency derivative $\dot\nu$ relative to the mean rotation
parameters for the model 1991--2009 from Table~\ref{model}
are presented in Figures~\ref{resid2}(c) and \ref{resid2}(d).
The plotted values of $\nu$ and $\dot{\nu}$ were
calculated from the local fits, performed to arrival time data
over the intervals of $\sim$ 150 days that overlapped by 75 days.
Then the eight points in $\Delta\nu$ will correspond
to a cycle duration of 600 days in the timing residuals.

It is clearly seen from Figure~\ref{resid2}(c) that the rotation frequency
of the pulsar B0919+06 slowly oscillates during the time interval
observed. In each successive cycle, the stage of increase in $\Delta\nu$
is followed by the stage of decrease in $\Delta\nu$.
This oscillatory process looks like a continuous sequence of
12 glitch-like events. In contrast to normal glitches, these events
can be classified as peculiar or slow glitches because they exhibit
long rise times of about a few hundred days. The similar glitch
phenomenon was observed earlier in the pulsar B1642$-$03 (see SH09).
Figure~\ref{resid2}(d) shows that the time changes in the frequency
derivative $\dot\nu$ are the effect of the changes in $\Delta\nu$.
The peaks of $\Delta\dot{\nu}$ define the steepness
of the front in the $\Delta\nu$ cycles.
In order to study the properties of the slow glitches observed, it
is necessary to obtain the shape of the $\Delta\nu$ cycles which
is not distorted by the presence of the timing noise.

\subsection{The Properties of the Slow Glitches Observed}

From Figure~\ref{resid2}(c), it is seen that the $\Delta\nu$ curve
shows appreciable deviations relative to the timing model
1991-2009. These deviations are a result of the variations in
$\dot\nu$ due to higher-order frequency derivatives which are not
taken into account by the timing model. In order to obtain the
undistorted shape of the $\Delta\nu$ cycles, we should calculate
the frequency residuals $\Delta\nu$ relative to several timing
models that describe the data over the shorter time intervals,
including a few slow glitches. A measured magnitude of $\ddot\nu$
is so large that a simple $\nu,\,\dot\nu$ spin down model cannot
be used to describe the experimental data even over the short
intervals. Besides, $\ddot\nu$ changes a sign within the interval
1991--2009. The positive value $\ddot\nu = 1.6\times 10^{-25}
s^{-3}$ measured over the interval 1991--2002 becomes negative
$\ddot\nu = -4.1\times 10^{-25} s^{-3}$ over the next interval
2002--2009. The maximum length of each short time interval was
determined by such quantity of the $\Delta\nu$ cycles that the
timing residuals relative to the model, including the mean values
of $\nu$, $\dot\nu$, and $\ddot\nu$ defined over this interval,
were nearly symmetrical with respect to the $X$-axis.

We found the three suitable time intervals on which the pulse arrival
times, corresponding to a few slow glitches, are well described by
the $\nu,\,\dot\nu,\ddot\nu$ model.
The timing model 1991-1995 was used to define the frequency residuals
$\Delta\nu$ for glitches 1, 2, 3, and 4; the model 1995-2004 -- for
glitches 5, 6, 7, and 8; the model 2004--2009 -- for glitches 9, 10,
11, and 12. The frequency residuals $\Delta\nu$ defined relative to
these three models were combined then into a single data set.
This adjusted set of the $\Delta\nu$ cycles describing the shape of
12 slow glitches is plotted in Figure~\ref{form}(a). It is seen
that the $\Delta\nu$ curve is nearly symmetrical with respect
to the zero line. This indicates that the shapes of the slow
glitches were reconstructed rather correctly
(compare with Figure~\ref{resid2}(c)).

The parameters describing the sequence of 12 slow glitches plotted
in Figure~\ref{form}(a) are given in Table~\ref{param}. The
parameters are presented in the following order: the glitch
number; epoch of the point $T_{min}$, which corresponds to the
minimum deviation of $\Delta{\nu_{min}}$; epoch of the point
$T_{max}$, which corresponds to the maximum deviation of
$\Delta{\nu_{max}}$; the glitch amplitude $\Delta{\nu_{g}}=
\Delta{\nu_{max}} +|\Delta{\nu_{min}}|$; the time interval to the
next glitch $\Delta{T_{max}}$; the time interval between the
minimal points of the $\Delta\nu$ cycles $\Delta{T_{min}}$;
the rise time interval $\Delta{T_{ris}} = T_{max} - T_{min}$;
the relaxation time interval after the glitch
$\Delta{T_{rel}} = T_{min} - T_{max}$.

Figure~\ref{form}(a) and Table~\ref{param} allow us to study the
properties of the slow glitches in the dependence on the glitch
number. The results are presented in Figure~\ref{depend}. The
top panel of Figure~\ref{depend} shows the relation between the
glitch amplitude and the glitch number. It is seen that all the
slow glitches have approximately an identical amplitude equal to
$3.5(0.5)\times 10^{-9}$ Hz, where the error in parentheses is the
formal standard deviation. The middle panel shows the relation
between the time intervals $\Delta{T_{max}}$ (the interval between
the successive glitches) and $\Delta{T_{min}}$ (the interval between
the minimal points of the $\Delta\nu$ cycles) and the glitch
number. This plot shows that the indicated intervals are
approximately equal to 580 and 600 days, respectively. The bottom
panel shows the relation between the glitch parameters
$\Delta{T_{rel}}$ (the relaxation time interval) and
$\Delta{T_{ris}}$ (the rise time interval) and the glitch number
and defines the average values of these parameters equal to 400
and 180 days, respectively.

These three relations indicate that all the slow glitches
observed have similar properties which can be described by
the following average parameters. The glitches have a small
absolute amplitude equal to $3.5\times 10^{-9}$ Hz. They are
characterized by the identical inter-glitch intervals
$\Delta{T_{max}}$ and approximately the same width of the
intervals $\Delta{T_{min}}$, equal to $\sim$ 600 days.
The glitches have similar signature related to a slow increase
in the rotation frequency during $\sim$ 200 days and the
subsequent relaxation back to the pre-glitch value during $\sim$
400 days. The relaxation after all the glitches can be described
by a linear curve as is seen from Figure~\ref{form}(a). These
properties suggest that the sequence of the slow glitches observed
can be approximated by a sawtooth-like function.

We created the model sawtooth-like curve using the indicated average
parameters. This model curve has the starting point MJD 48350 and
includes 10 cycles consisting of two stages -- the stage of a linear
glitch arising with a timescale of 200 days and the stage
of a linear post-glitch relaxation with a timescale of 400 days.
Only two glitches observed, 2 and 3, take off from this sequence.
An analysis of the $\Delta\nu$ cycles showed that event 2
represents the sum of two partially overlapping glitches 2 and 3.
Glitch 3 defines the starting point of a new phase in the sequence
of the slow glitches. Therefore, event 2 should be described by
three stages -- the stage of a linear glitch arising with
a timescale of 200 days is followed by the stage in which
the glitch amplitude keeps constant within 400 days (the duration
of this interval corresponds to the duration of the relaxation time
interval $\Delta{T_{rel}}$) and only then is followed by a linear
post-glitch relaxation with a timescale of 400 days.
The derived values for this model sawtooth-like curve are given
in Table~\ref{sawt}.

Figure~\ref{form}(b) shows a model sawtooth-like curve with
a period of 600 days which is superimposed on the glitch sequence
observed. It is seen that the maxima of the model curve well
coincide with the maxima of nearly all the slow glitches. Only
the maxima of
glitches 8 and 9 slightly do not correspond to the model curve.
However, as is seen from the plot, the slight deviations of the
amplitude and phase of these cycles from a model curve do not
change a phase of the next cycles of the sequence. Probably, the
shape of these glitches was not restored precisely. The model
curve very well describes partially overlapping glitches 2 and
3. It is seen that in this range there was a phase shift for 400 days,
exactly equal to $\Delta{T_{rel}}$. After that, point 3
started marking the starting point of a new phase in the sequence
of the slow glitches. Despite the phase shift between points 2
and 3, we suppose that the model sawtooth-like function is a periodic
function.
A comparison of the model parameters $T_{min}$ and $T_{max}$, given in
Table~\ref{sawt}, and the same experimental parameters, listed in
Table~\ref{param}, shows that these parameters agree well within
a precision limited by the time resolution of $\sim$ 100 days.
We conclude that a model periodic sawtooth-like function approximates
a sequence of the slow glitches observed very well. This result
indicates a surprising regularity in the occurrence of slow
glitches -- the time intervals between the slow glitches, equal to
600 days, keep constant within approximately of 100 days
during $\sim$ 19 years.

Thus, we found that the cause of slow oscillations in the rotation
frequency of PSR B0919+06 over the 1991--2009 interval lies in a
continuous generation of slow glitches. The originality of these
slow glitches is that they have similar properties and their
sequence is described by a periodic sawtooth-like function. Estimates
show that the time intervals between glitches remain constant within
100 days during $\sim$ 19 years. The regular occurrence of similar
slow glitches creates the sawtooth-like pattern in the frequency
residuals with a period of 600 days. The presence of
the regular pattern over a long time-span of 19 years indicates
that the slow glitches observed are not random events.

This conclusion is also valid for the full time span 1979--2009.
Figure~\ref{resid1}(c) displays the curve B which represents the
model sawtooth-like curve extended by an initial data segment
1979--1986. Poor and irregular experimental points obtained over
this interval give an approximate picture of oscillations
observed in the frequency residuals $\Delta\nu$ and the timing
residuals. Nevertheless, as is seen in Figures~\ref{resid1}(b) and
~\ref{resid1}(c), the cycles in the model sawtooth-like curve B
correspond well to cyclical changes in the frequency and
timing residuals. The curve B shows how cyclical changes
in the rotation frequency could look if the pulsar's rotation
was described by a simple spin-down model over the interval
1979--2009.

\section{Summary}
An analysis of the rotation behavior of the pulsar B0919+06 over
the 30-year period from 1979 to 2009 has shown that the nature of
cyclical changes in the timing residuals from this pulsar implies
a continuous generation of slow glitches which have similar properties.
They are characterized by small sizes of $3.5\times10^{-9}$ Hz,
long rise times of 200 days, and relaxation time intervals of 400
days. We found that a sequence of the slow glitches observed over
the 1991-2009 interval is well approximated by a periodic
sawtooth-like function with a period of 600 days.
Estimates show that the time intervals between the slow glitches
observed keep constant within approximately 100 days
during this period. In Figure~\ref{resid1}(c), the curve B suggests
that this conclusion can be generalized to all the observational
interval 1979--2009. We may suppose that a sequence of similar
slow glitches occurring at regular time intervals produced
a sawtooth-like modulation of the rotation frequency, superimposed
on the secular spin down, during all the period 1979--2009.
An analysis of the pulse arrival times showed that the process,
responsible for a generation of the slow glitches over $\sim$ 30
years, was interrupted by the large glitch of magnitude
$\Delta\nu/{\nu}=1.3\times 10^{-6}$ that occurred in 2009 November 5.

\section{Discussion}

Slow glitches as a unique glitch phenomenon were originally
detected for the pulsar B1822$-$09 in the timing observations
which are carried out at the Pushchino radio telescope since 1991
\citep{sha98}. Further observations showed that this pulsar
experienced a series of five slow glitches and also suffered three
glitches of normal signature
\citep{sha00,zou04,sha05,sha07,sha09a,yua10}. The slow glitches
observed were characterized by a gradual increase in the rotation
frequency with a long rise time of 100--300 days. It was found
that all these slow glitches were the components of one process
that acted continuously during $\sim$ 10 years from 1995 to 2004.
Then these events were followed by two glitches of normal signature.
The glitches of magnitude $\Delta\nu/{\nu}=6.7\times 10^{-9}$ and
$\Delta\nu/{\nu}=121\times 10^{-9}$ occurred in 2006 January
and 2007 January, respectively \citep{sha09a}. All these events
clearly show the presence of two types of discontinuities in the
rotation frequency of the pulsar B1822$-$09. These discontinuities
can be classified as normal and slow glitches.

The pulsar B1642$-$03 is the second pulsar known, after PSR
B1822-09, in the rotation frequency of which slow glitches were
revealed. As shown in SH09, the rotation frequency of this pulsar
undergoes continuous generation of slow glitches and over the
40-year period of observations this pulsar suffered eight slow
glitches. The amplitude of these glitches and the time interval to
the next glitch obey a clear linear relation. This dependence
gives strong evidence against the statement that slow glitches can
be caused by the same process as the timing noise in pulsars
\citep{hob10}. The existence of the modulation process which
causes the discrete changes of the glitch amplitudes and the
post-glitch time intervals in the dependence on the serial number
of the glitch in a given modulation period also confirms that the
slow glitches observed are the distinct events. The indicated
dependencies allow us to predict the epochs and sizes of new
glitches in this pulsar. The verification of these predictions can
be obtained in the near future, in 2013. This verification
will provide strongest evidence that the slow glitches observed
are a unique glitch phenomenon.

\begin{figure}
\epsscale{.70}
\plotone{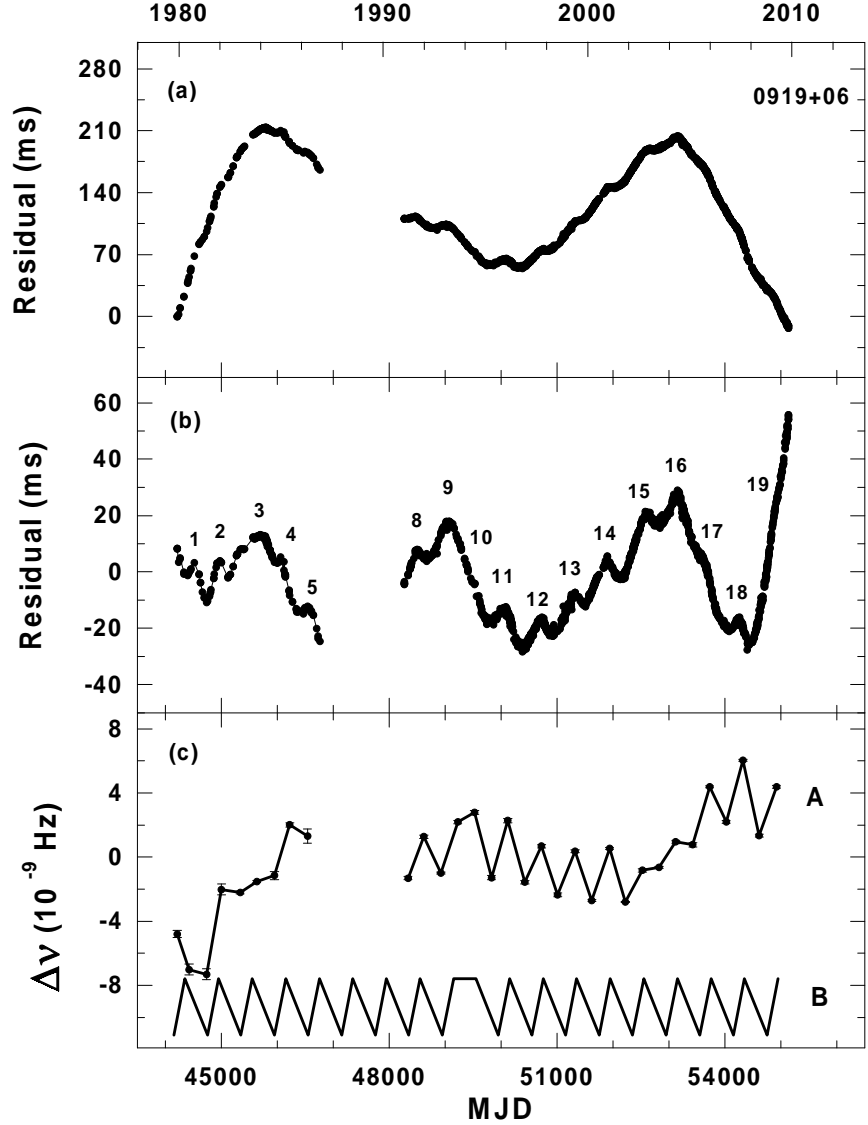}
 \caption{Timing behavior of PSR B0919+06 over the 30-year
     interval from 1979 to 2009. The four-year data gap is seen
     between 1987 and 1990. (a) The timing residuals
     after subtraction of the polynomial for $\nu$, $\dot\nu$,
     and $\ddot\nu$ for all the pulse arrival times. The residual
     curve shows the slight short-term cyclical structure. (b) The timing
     residuals after subtraction of the polynomial for $\nu$ and
     three frequency derivatives. The short-term cyclical structure
     contains 19 cycles with spacing of maxima of about 600 days.
     (c) The curve A displays the frequency residuals $\Delta\nu$
     relative to the timing model 1979--2009. It is seen that
     the signature of the $\Delta\nu$ changes has a sawtooth-like
     character. The curve B is the model sawtooth-like curve
     that shows how cyclical changes in the rotation frequency
     could look if the pulsar's rotation was described by a simple
     spin-down model over the 1979-2009 interval.
     \label{resid1}}
\end{figure}

\begin{figure}
\epsscale{.70}
\plotone{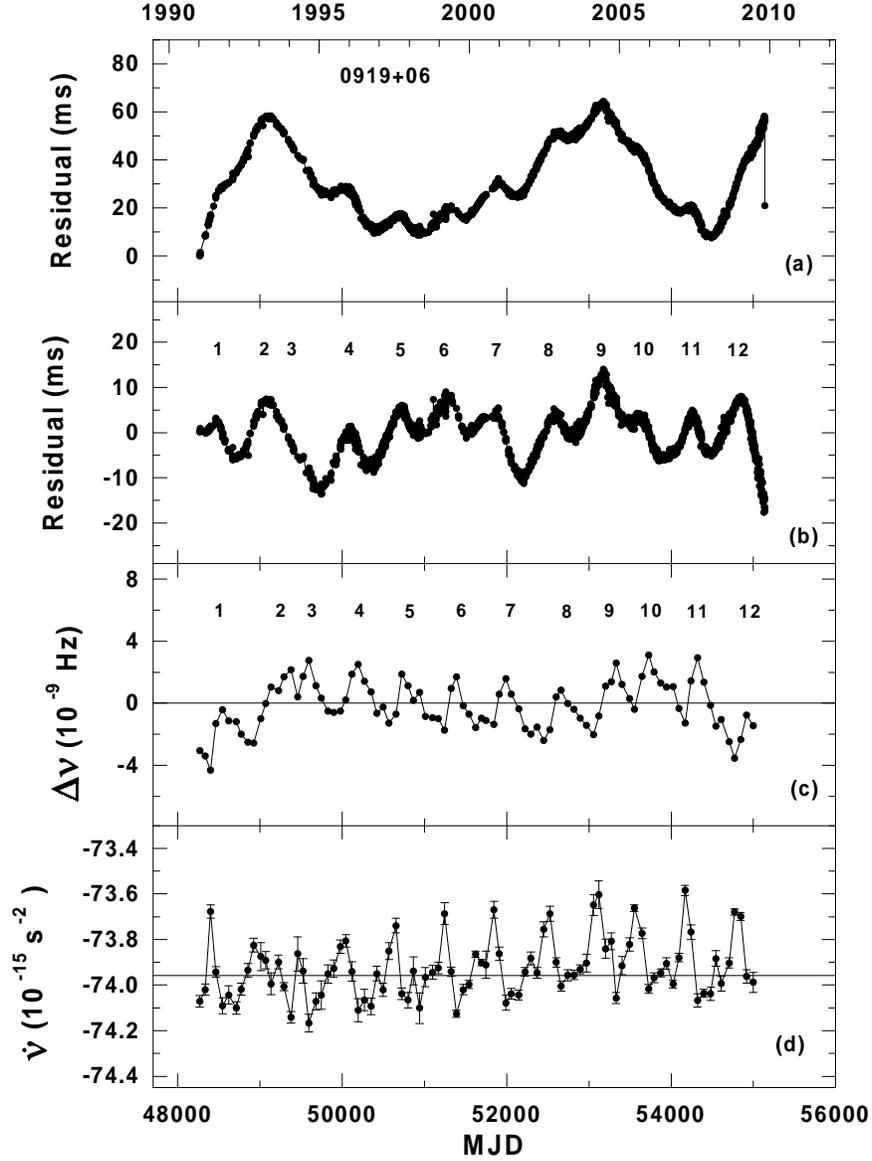}
 \caption{Timing behavior of PSR B0919+06 between 1991 and 2009.
     (a) The timing residuals after subtraction of the polynomial
     for $\nu$ and two frequency derivatives. The last point in the
     residual curve indicates the large glitch of 2009 November.
     The slight cyclical structure is clearly seen in the residual
     curve. (b) The timing residuals after subtraction of
     the polynomial for $\nu$ and five frequency
     derivatives. The clear cyclical structure contains 12 cycles
     with the amplitudes of $\sim$ 10 ms and the spacing of maxima
     of $\sim$ 600 days. (c) The frequency residuals relative to the
     timing model 1991--2009. Slow oscillations in $\Delta\nu$ look
     like a continuous sequence of 12 glitch-like events. (d) The changes
     in the frequency first derivative $\dot\nu$ with time. The peaks
     of $\Delta\dot{\nu}$ define the steepness of the front in $\Delta\nu$.
     \label{resid2}}
\end{figure}

\begin{figure}
\epsscale{.80}
\plotone{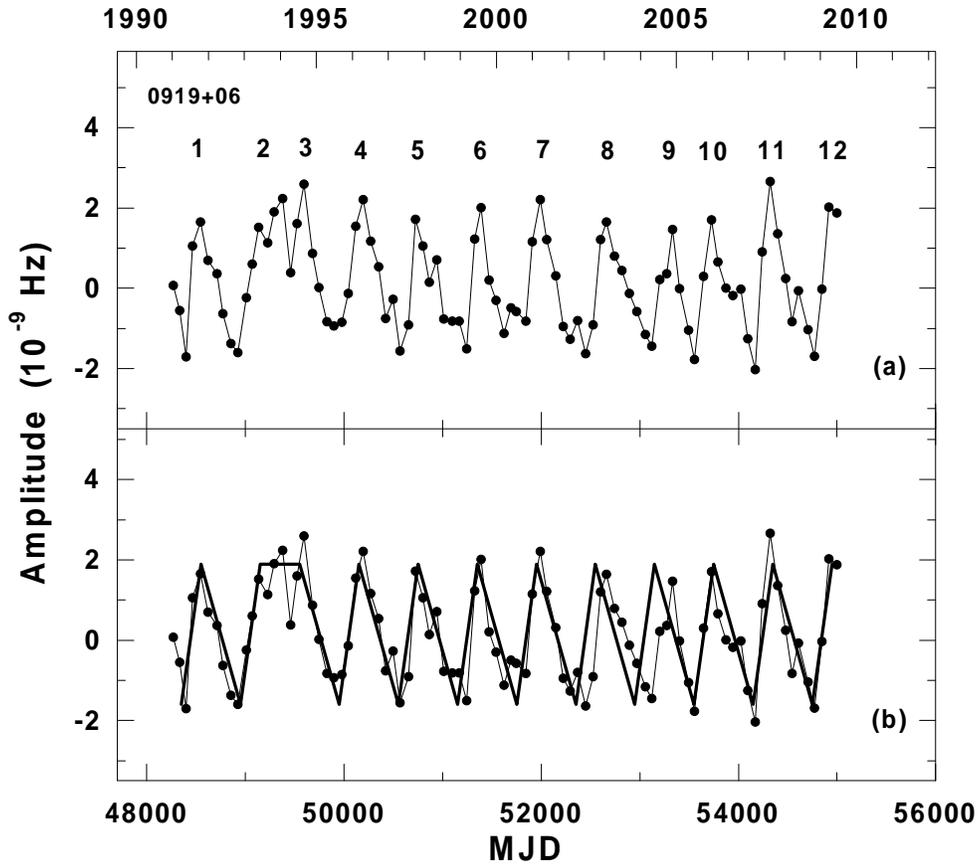}
\caption{Shapes of 12 successive slow glitches observed over the interval
      1991--2009. (a) The adjusted set of the $\Delta\nu$ cycles describing
      the shapes of 12 successive slow glitches. (b) The model
      sawtooth-like curve with a period of 600 days (marked by the bold
      line) is superimposed on a sequence of the slow glitches,
      plotted as in the top panel (a). It is seen that
      the model sawtooth-like function approximates very well
      a sequence of the slow glitches observed.
      \label{form}}
\end{figure}

\begin{figure}
\epsscale{.70}
\plotone{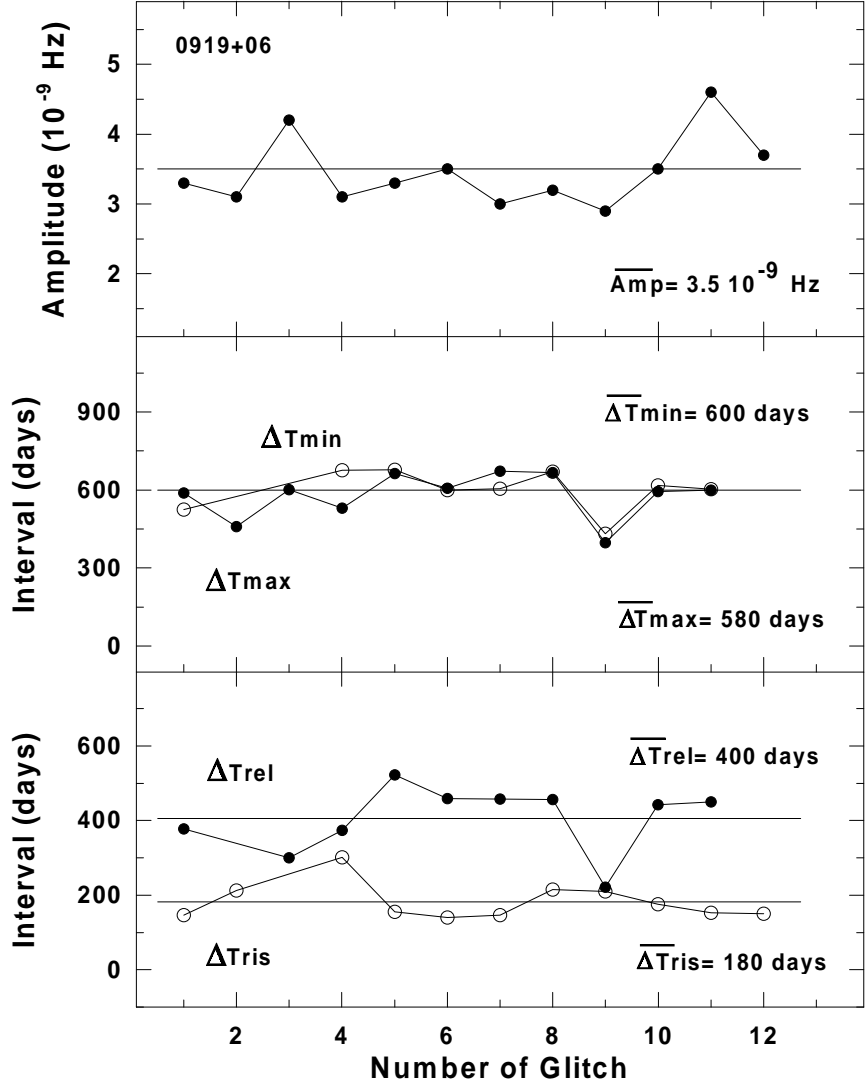}
\caption{Relation between the parameters of 12 slow glitches and
    the glitch number. The top panel presents a relation between
    the glitch amplitude and the glitch number. All the slow glitches
    have approximately an identical amplitude. Its average magnitude
    equals $\sim 3.5\times10^{-9}$ Hz. The mean panel presents
    a relation between the time interval $\Delta{T_{max}}$
    between glitches and the time interval $\Delta{T_{min}}$
    between the minimal points of the $\Delta\nu$ cycles and
    the glitch number. Their average magnitudes equal $\sim$ 580
    and $\sim$ 600 days, respectively.
    The bottom panel presents a relation between the relaxation time
    interval $\Delta{T_{rel}}$ and the rise time interval
    $\Delta{T_{ris}}$ and the glitch number. Their average magnitudes
    equal $\sim$ 400 and $\sim$ 180 days, respectively.
    The derived average parameters indicate that all the slow glitches
    have similar properties.
     \label{depend}}
\end{figure}

The pulsar B0919+06 is the third pulsar in our research that
experienced slow glitches in its rotation frequency.
The main properties of slow glitches observed are a similarity in
their signatures and a regularity in their occurrence.
The regular occurrence of similar slow glitches produces a periodic
sawtooth-like modulation of the rotation frequency with a period of
600 days. As discussed above, the timing residuals do not show strictly
periodic cyclical changes because this pulsar possesses a high level
of timing noise and the observed amplitude and the shape of
the cycles depend on a polynomial model fitted. A derived
sawtooth-like modulation of the rotation frequency, shown in
Figure~\ref{resid1}(c) by the curve B, is a result of reconstruction
of cyclical changes that could take place in the pulsar's rotation
frequency if the pulsar rotation is modeled by a simple spin down model.
The curve B reflects the properties of an actual process that generated
slow glitches during $\sim$ 30 years. This process was interrupted
by a large glitch that occurred in 2009 November.
The pulsar B0919+06, as well as B1822$-$09, clearly exhibits that
the pulsar's rotation frequency underwent two types of
discontinuities that can be classified as normal and slow glitches.

Comparison of the rotation parameters for the pulsars showing the slow
glitches is presented in Table~\ref{3psr} where the pulsars are listed
in order of decreasing of their age. The table gives the pulsar's B1950
name, the rotation parameters $\nu$, $\dot\nu$, and $\ddot\nu$,
characteristic age $\tau = {P}/{2{\dot{P}}}$, and  surface magnetic
field $B=3.2\times10^{19}({P}{\dot{P}})^{1/2}$ G. It is seen that
all these pulsars are relatively old pulsars with ages greater
than $\sim 10^{5}$ years. The oldest pulsar B1642$-$03  has only
the slow glitches. Their amplitude and the time interval between
glitches strictly obey a certain law, that is, the glitch
sequence in this pulsar possesses the predicted, steady-state
properties. The two others, B0919+06 and B1822$-$09, are substantially
younger, have the higher frequency derivatives $\dot\nu$ and
$\ddot\nu$, and the stronger magnetic fields. They experience
large glitches of normal signature that followed the oscillatory
process in the rotation frequency identified with slow glitches.
A comparison of the pulsar parameters in this table indicates that
a tendency to have a sequence of slow glitches with steady-state
properties is correlated with the characteristic age of the pulsar.
Probably, the number of pulsars with slow glitches in their
rotation frequency will considerably increase in the future.
The candidates can be pulsars that exhibit cyclical changes in
the timing residuals.

Long sequences of the timing residuals were recently published for
366 pulsars observed at the Jodrell Bank between 1968 and 2006
\citep{hob10}. A detailed analysis of these data shows that the timing
residuals of some pulsars have clear cyclical changes during all
the period of observations. According to \citet{hob10},
the quasi-periodic structure in the timing residuals is clearly
visible for six pulsars: B1540$-$06, B1642$-$03, B1818$-$04,
B1826$-$17, B1828$-$11, and B2148+63. It should be noted that all
these pulsars are old pulsars with ages varying from $1\times 10^{5}$
to $3.5\times 10^{7}$ years. Among these pulsars there are two pulsars
B1642$-$03 and B1828$-$11 that were investigated earlier. The results of
spectral analysis by \citet{hob10} for these two pulsars agree with
the results of previous papers \citep{sta00,sha01}. It is known that
clear periodic structure in the timing residuals of B1828$-$11 that
is accompanied by correlated pulse shape changes is explained
by free-precession of the neutron star \citep{sta00}.
In the case of PSR 1642$-$03, multiple low frequency components in
the power spectrum of timing residuals can be explained well as
a result of continuous generation of slow glitches, the amplitude
of which is correlated with the time interval following
the glitch \citep{sha09b}.

As reported by \citet{yua10}, the slow glitches have been identified
yet for two pulsars J0631+1036 and B1907+10. The first pulsar
is a young pulsar with $\tau \sim 4.4\times 10^{4}$ years, and the second
one is substantially older with $\tau \sim 1.7\times 10^{6}$ years.
The latter has cyclical timing residuals, as is shown in Figure 3
of \citet{hob10}, that can imply cyclical changes in
the rotation frequency.

It is thought that pulsar glitches reflect a variable coupling
between the solid crust of a neutron star and the superfluid
interior rotating more rapidly than the solid crust. In terms of
vortex pinning models, the origin of glitches can be explained by
the catastrophic unpinning of superfluid vortices
\citep{and75,alp84,alp89,alp93,pin85}. These models provide
a satisfactory interpretation of large glitches in pulsars.

Now the interpretation of a phenomenon of slow glitches is uncertain.
According to \citet{hob10}, the slow glitch phenomenon is
a quasi-periodic component of timing noise, unrelated to normal
glitches. At the same time, the slow glitches revealed in PSR B1642$-$03
possess the properties which meet the requirements of the glitch
models: 1) all the slow glitches observed have significant
exponential decay after glitch which is characterized by the same
parameter $Q\sim0.9$, 2) the size of glitches and the time
interval to the following glitch obey a strong linear relation.
The third property, the presence of a modulation process which
forces the glitch amplitudes and the inter-glitch intervals to
change with a discrete step, was not considered yet by any theory
of pulsar glitches. In the case of slow glitches, it is necessary
to account for the cause of a continuous generation of slow glitches.

For PSR B0919+06, a sawtooth-like modulation of the rotation
frequency with a period of 600 days could be interpreted
by the free precession of an isolated pulsar if this modulation was
accompanied by correlated observable changes in the pulse profile
\citep{sha77,nel90,cor93,sta00}. We detected no pulse profile
changes during our observations at 112 MHz.
A precession model requires a strictly periodic modulation of
the pulsar's rotation frequency. The presence of the phase shift,
equal to 400 days, in the modulation curve B between cycles 9 and 10,
as is shown in Figure~\ref{resid1}(c), contradicts this requirement.
These two arguments testify against the interpretation of this
sawtooth-like modulation in the terms of a free precession model.
Besides, this pulsar has experienced a large glitch.
As discussed by \citet{lin07}, a slowly precessing neutron star
cannot produce a glitch. The useful information for explanation of an
origin of slow glitches can be obtained from further timing
observations of PSR B0919+06. The observations in the nearest
5--10 years allow us to find out whether there is a relationship
between different phenomena observed in this pulsar, such as a large
glitch and a sequence of the slow glitches. Whether a large glitch
has put a stop to a process of generation of slow glitches or this
process continues to work after a large glitch in the same mode.

\acknowledgments I should like to thank R. D. Dagkesamansky
for useful discussion and comments, the staff of the PRAO for their
aid in carrying out the many-year observations of this pulsar on
the LPA antenna. This work was supported by the European Commission
6th Framework Program, Square Kilometre Array Design Studies
(SKADS project, contract no. 011938) and the Russian Foundation
for Basic Research (grant 09-02-00473).

\newpage
\clearpage

\newpage
\clearpage
\newpage
\clearpage
\newpage
\clearpage

\newpage
\clearpage

\clearpage
\begin{deluxetable}{lc}
\tablecaption{The Glitch Parameters for PSR B0919+06
        \label{glitchp}}
\tablewidth{0pt}
\tablehead{
\colhead{Glitch Parameters} & \colhead{Values}
}
\startdata
 Pre-glitch parameters            &                   \\
 MJD range                        & 54892--55139      \\
 $\nu$ (Hz)                       & 2.32219343204(6)  \\
 $\dot\nu$ ($10^{-15}\,s^{-2}$)   & -73.931(6)        \\
 Epoch (MJD)                      & 54892.8436        \\
 RMS Timing Residual (ms)         & 0.6               \\
\tableline
 Post-glitch parameters           &                   \\
 MJD range                        & 55140--55254      \\
 $\nu$ (Hz)                       & 2.3221947716(7)   \\
 $\dot\nu$ ($10^{-15}\,s^{-2}$)   & -73.43(8)         \\
 Epoch (MJD)                      & 55140.1604        \\
 RMS Timing Residual (ms)         & 0.9               \\
\tableline
 Glitch parameters                      &            \\
 ${\Delta\nu}/{\nu}\,(10^{-9})$         & 1257.1(3)  \\
 ${\Delta\dot\nu}/{\dot\nu}\,(10^{-3})$ & -7(1)      \\
 Epoch (MJD)                            & 55139.8(1) \\
\enddata
\end{deluxetable}

\clearpage
\begin{deluxetable}{lcc}
\tablecaption{The Rotation Parameters for PSR B0919+06
              \label{model}}
\tablewidth{0pt}
\tablehead{
\colhead{Parameter} & \colhead{Model 1979--2009} &
\colhead{Model 1991--2009}\\
}
\startdata
 MJD range                & 44210--55139       & 48267--55139   \\
 $\nu$ (Hz)               &2.32226169326(9)  &2.32223575479(4)  \\
 $\dot\nu$ ($10^{-15}\,s^{-2}$)   &-74.0527(5) &-74.0211(4)     \\
 $\ddot\nu$ ($10^{-25}\,s^{-3}$)  & 1.908(13)  & 3.245(12)      \\
 Epoch (MJD)              & 44210.5817         & 48266.9832     \\
 RMS Timing Residual (ms) & 57                 & 15             \\
\enddata
\end{deluxetable}

\clearpage
\begin{deluxetable}{lccccccccc}
\tabletypesize{\scriptsize}
\tablecaption{The Observed Parameters for a Sequence of 12 Slow
       Glitches Revealed in PSR B0919+06 Between 1991 and 2009
       (see Figure~\ref{form}(a))
       \label{param}}
\tablewidth{0pt}
\tablehead{
\colhead{No.} & \colhead{${T_{min}}$} & \colhead{$\Delta{\nu_{min}}$} &
\colhead{${T_{max}}$} & \colhead{$\Delta{\nu_{max}}$} &
\colhead{$\Delta{\nu_{g}}$} &\colhead{$\Delta{T_{max}}$} &
\colhead{$\Delta{T_{min}}$} &\colhead{$\Delta{T_{ris}}$} &
\colhead{$\Delta{T_{rel}}$}\\
\colhead{} & \colhead{(MJD)} & \colhead{($10^{-9}$ Hz)} & \colhead{(MJD)} &
\colhead{($10^{-9}$ Hz)} & \colhead{($10^{-9}$ Hz)} &
\colhead{(days)} & \colhead{(days)} &\colhead{(days)} & \colhead{(days)}
}
\startdata
 1   &48398 &-1.7 &48545  & 1.6  &3.3  &589  & 524    &147  & 377  \\
 2   &48922 &-1.6 &49134  & 1.5  &3.1  &459  &  --    &212  & --   \\
 3   & --   & --  &49593  & 2.6  &4.2  &601  &  --    & --  & 300  \\
 4   &49893 &-0.9 &50194  & 2.2  &3.1  &529  & 675    &301  & 374  \\
 5   &50568 &-1.6 &50723  & 1.7  &3.3  &663  & 678    &155  & 523  \\
 6   &51246 &-1.5 &51386  & 2.0  &3.5  &606  & 599    &140  & 459  \\
 7   &51845 &-0.8 &51992  & 2.2  &3.0  &672  & 604    &147  & 457  \\
 8   &52449 &-1.6 &52664  & 1.6  &3.2  &666  & 671    &215  & 456  \\
 9   &53120 &-1.4 &53330  & 1.5  &2.9  &396  & 431    &210  & 221  \\
 10  &53551 &-1.8 &53726  & 1.7  &3.5  &594  & 617    &175  & 442  \\
 11  &54168 &-2.0 &54320  & 2.6  &4.6  &600  & 602    &152  & 450  \\
 12  &54770 &-1.7 &54920  & 2.0  &3.7  & --  &  --    &150  & --   \\
\enddata
\tablecomments{In column order, the table gives the glitch number;
    epoch of the point $T_{min}$, which corresponds to
    the minimum deviation of $\Delta{\nu_{min}}$; epoch of
    the point $T_{max}$, which corresponds to the maximum
    deviation of $\Delta{\nu_{max}}$; the glitch amplitude
    $\Delta{\nu_{g}}= \Delta{\nu_{max}} +|\Delta{\nu_{min}}|$;
    the time interval to the next glitch $\Delta{T_{max}}$;
    the time interval between the minimal points of the $\Delta\nu$
    cycle $\Delta{T_{min}}$;
    the rise time interval $\Delta{T_{ris}} = T_{max} - T_{min}$;
    the relaxation time interval after the glitch
    $\Delta{T_{rel}} = T_{min} - T_{max}$.}
\end{deluxetable}
\clearpage
\begin{deluxetable}{lcccc}
\tabletypesize{\scriptsize}
\tablecaption{The Parameters for a Model Sawtooth-Like Curve
       Approximating a Sequence of 12 Slow Glitches Observed
       (see Figure~\ref{form}(b))
     \label{sawt}}
\tablewidth{0pt}
\tablehead{
\colhead{No.} & \colhead{${T_{min}}$} & \colhead{$\Delta{\nu_{min}}$} &
\colhead{${T_{max}}$} & \colhead{$\Delta{\nu_{max}}$} \\
\colhead{} & \colhead{(MJD)} & \colhead{($10^{-9}$ Hz)} & \colhead{(MJD)} &
\colhead{($10^{-9}$ Hz)}
}
\startdata
 1   &48350 &-1.6 &48550  & 1.9     \\
 2   &48950 &-1.6 &49150  & 1.9     \\
 3   & --   & --  &49550  & 1.9     \\
 4   &49950 &-1.6 &50150  & 1.9     \\
 5   &50550 &-1.6 &50750  & 1.9     \\
 6   &51150 &-1.6 &51350  & 1.9     \\
 7   &51750 &-1.6 &51950  & 1.9     \\
 8   &52350 &-1.6 &52550  & 1.9     \\
 9   &52950 &-1.6 &53150  & 1.9     \\
 10  &53550 &-1.6 &53750  & 1.9     \\
 11  &54150 &-1.6 &54350  & 1.9     \\
 12  &54750 &-1.6 &54950  & 1.9     \\
\enddata
\tablecomments{In column order, the table gives the glitch number;
    epochs of the points $T_{min}$, which correspond to
    the minimum deviations of $\Delta{\nu_{min}}$; epochs of
    the points $T_{max}$, which correspond to the maximum
    deviations of $\Delta{\nu_{max}}$. The glitch amplitude
    equals $3.5\times10^{-9}$ Hz, $\Delta{T_{ris}}=200$ days,
    and $\Delta{T_{rel}}=400$ days.}
\end{deluxetable}
\clearpage
\begin{deluxetable}{lccccc}
\tablecaption{The Parameters for the Pulsars Showing Slow Glitches
       \label{3psr}}
\tablewidth{0pt}
\tablehead{
\colhead{PSR} & \colhead{$\nu$} & \colhead{$\dot\nu$} &
\colhead{$\ddot\nu$} & \colhead{$\tau$} & \colhead{B}   \\
\colhead{} & \colhead{$(s^{-1})$} & \colhead{$(10^{-15} s^{-2})$}&
\colhead{$(10^{-25} s^{-3})$} & \colhead{(years)} & \colhead{$(G)$}
}
\startdata
 B1642$-$03 &2.579 &-11.84 &0.02  & $3.4\times10^{6}$  &$8.3\times10^{11}$  \\
 B0919+06   &2.322 &-74.05 &1.90  & $5.0\times10^{5}$  &$2.5\times10^{12}$  \\
 B1822$-$09 &1.300 &-88.59 &9.00  & $2.3\times10^{5}$  &$6.4\times10^{12}$  \\
\enddata
\end{deluxetable}

\end{document}